\documentclass[doublecol]{epl2} 
\usepackage{amsmath,graphicx}
\title{On the role of conserved moieties in shaping the robustness and production capabilities of reaction networks}

\shorttitle{Title} 

\author{Andrea De Martino\inst{1} \and Carlotta Martelli\inst{2}\footnote{Present address: Department of Molecular, Cellular and Developmental Biology, Yale University, KBT 1048, P.O. Box 208103
New Haven, CT 06520 (USA)} \and Francesco A. Massucci\inst{2}\footnote{Present address: Department of Mathematics, King's College London, Strand, London WC2R 2LS (UK)}}
\shortauthor{A. De Martino, C. Martelli and F. A. Massucci}

\institute{                    
  \inst{1} CNR/INFM (SMC), Dipartimento di Fisica, Sapienza Universit\`a di Roma, p.le A. Moro 2, 00185 Roma (Italy)\\
  \inst{2} Dipartimento di Fisica, Sapienza Universit\`a di Roma, p.le A. Moro 2, 00185 Roma (Italy)
}

\abstract{We study a simplified, solvable model of a fully-connected metabolic network with constrained quenched disorder to mimic the conservation laws imposed by stoichiometry on chemical reactions. Within a spin-glass type of approach, we show that in presence of a conserved metabolic pool the flux state corresponding to maximal growth is stationary independently of the pool size. In addition, and at odds with the case of unconstrained networks, the volume of optimal flux configurations remains finite, indicating that the frustration imposed by stoichiometric constraints, while reducing growth capabilities, confers robustness and flexibility to the system. These results have a clear biological interpretation and provide a basic, fully analytical explanation to features recently observed in real metabolic networks.}

\begin{document}

\maketitle

Metabolic networks represent the biochemical machinery by which cells dispose of the nutrients found in their surrounding environment to produce the macromolecules needed for their survival, including nucleic acids, membranes, cell walls, proteins and the energy carrier ATP. Understanding how functionality emerges from such a complex system of biochemical reactions is a major issue with important implications in bioengineering and pharmacology. Unfortunately, detailed kinetic models of genome-wide metabolic networks are computationally prohibitive and fine tuning of parameters would require an amount of biological information that is not available. It is therefore important to develop methods to predict the organization of metabolic fluxes from the sole knowledge of the stoichiometry, which represents the full available information about the network topology and about the proportions in which different reagents are interconnected as a set of input-output relations. This problem is central in systems biology. A key question in this context is whether the physiological states observed in real cells are optimal from a metabolic viewpoint, i.e. if reaction fluxes self-organize so as to maximize the production capabilities of the network (or of a smaller set of metabolites) \cite{pals,segre,bara,sauer}. 

It is quite intuitive that, in a generic situation where reagents flow in a heterogeneous input-output network, the ways in which fluxes can be chosen will be heavily constrained by the need to meet prescribed production objectives, while the dependence of the global properties on the specific realization of input and output coefficients will become weaker and weaker for larger networks. Ultimately, it is reasonable to think that the feasibility of flux configurations meeting all constraints, the robustness of the solutions against localized flux variations (or reaction knock-outs), and the productive efficiency of solutions will depend crucially on structural parameters like the ratio between the number of reactions (variables) and that of reagents (constraints). 

But in order to address the question of optimality in a realistic biochemical network it is necessary to take into account the peculiar nature of stoichiometric coefficients, which enforce mass balance conditions at each reaction node in the network. Such relations in turn define a family of conservation laws reflecting the existence of  dynamical invariants of purely topological origin that are able to affect the kinetics of the system as a whole \cite{fam}. To be explicit, let $a_i^\mu$ (respectively, $b_i^\mu$) denote the output (respectively, input) stoichiometric coefficient of metabolite $\mu\in\{1,\ldots, M\}$ in reaction $i\in\{1,\ldots, N\}$. A group $G$ of metabolites satisfying
\begin{equation}\label{cp}
\sum_{\mu\in G}(a_i^\mu-b_i^\mu)=0~~~~~\forall i
\end{equation}
is such that the total number of molecules in the pool, or the aggregate concentration, does not change in time. Such pools are ubiquitous in real metabolic networks: an important example is the total quantity of ATP and ADP (the `adenylate moiety') which remain constant during metabolic activity, as ATP is continuously discharged to ADP and vice-versa \cite{fam}.

In this Letter we study how the presence of conserved metabolic pools, reflecting the underlying mass-balance conditions of stoichiometric origin, affect the global growth capabilities and the volume of the solution space in a toy metabolic network where reagents are fully interconnected and stoichiometric coefficients (which play the role of a quenched disorder) satisfy laws such as (\ref{cp}). To be as general as possible and to highlight the ingredient of conserved pools, we adopt the framework of Von Neumann's expanding model \cite{vn}, under which one aims at computing the maximum value of $\rho>0$ for which the system of inequalities
\begin{equation}\label{vn}
\sum_{i=1}^N s_i(a_i^\mu-\rho b_i^\mu)\geq 0~~~~~\forall\mu
\end{equation}
admits solution(s) in the form of non-zero flux vectors $\boldsymbol{s}=\{s_i\geq 0\}$. From a physical point of view, the conditions (\ref{vn}) ensure that the total output of each metabolite $\mu$ is at least $\rho$ times the total input, so that if the maximum achievable $\rho$ (which we shall denote by $\rho^\star$) exceeds (respectively, is smaller than) one, the network is in an expanding (respectively, contracting) state. $\rho^\star=1$ signals instead a stationary network. (See \cite{fulc} for a more thorough description of Von Neumann's setup and a dynamical derivation of (\ref{vn}).) 

Following \cite{fulc}, we assume that every reaction produces and consumes (in different proportions) every metabolite. In such a fully-connected framework, when $N\to\infty$ and the input and output matrix are independent, identically distributed quenched random variables, it has been shown that $\rho^\star$ depends only on $n=N/M$ in such a way that the system undergoes a transition from a contracting to an expanding phase at $n_c=1$. Moreover, (\ref{vn}) admits a unique feasible flux configuation when $\rho=\rho^\star$. The advantage of using an unrealistic fully-connected setup lies mainly in its analytic tractability. We shall see that indeed the fully connected model provides an excellent proxy for a real metabolic network, at least as long as production capabilities and solution space are concerned. Graphical versions of the model yield essentially the same scenario \cite{graph}. 

To keep things mathematically simple, we account for conserved metabolic pools by constraining each disorder sample (i.e., each realization of the stoichiometric coefficients) to embed a pool formed by a finite fraction $\phi$ of metabolites. In other terms, we request that
\begin{equation}\label{con}
\sum_{\mu=1}^M z^\mu(a_i^\mu-b_i^\mu)=0~~~~~\forall i
\end{equation}
where $z^\mu$ is a quenched random variable that equals $1$ with probability $\phi$ and zero otherwise. In this way, we include a single conserved pool in the system. It will become clear that both the number of pools and their size do not affect the growth properties in the framework we consider.

For a start, note that multiplying each term in (\ref{vn}) by $z^\mu$ and summing over $\mu$ one easily obtains
\begin{equation}
\sum_{i=1}^N s_i \sum_{\mu=1}^M z^\mu(a_i^\mu-b_i^\mu)+(1-\rho)\sum_{i=1}^N s_i
\sum_{\mu=1}^M z^\mu b_i^\mu\geq 0
\end{equation}
If the $z^\mu$'s are such that (\ref{con}) holds, then either $\rho\leq 1$ or all $s_i$'s connected to a metabolite in the conserved pool must vanish. The latter condition however leads to the null solution $s_i=0~\forall i$ in a fully connected model and must be discarded. Hence necessarily $\rho\leq 1$. One then sees that an expanding regime cannot occur in a system with a conserved metabolic pool, independently of its size. This suggests a radically different and more realistic scenario than the unconstrained case. In addition, it is easy to understand that the inequalities (\ref{vn}) must become equalities at $\rho=1$ for all metabolites belonging to a conserved pool. In other words: if a metabolite belongs to a conserved pool there can be no net production or consumption of it in the optimal state. 

To get a more thorough insight, it is necessary to compute $\rho^\star$ exactly as a function of $n$. As in \cite{fulc}, the calculation  can be carried out in two steps. First, we compute the typical volume of flux configurations compatible with (\ref{vn}) and (\ref{con}). This requires the calculation of an average over the constrained quenched disorder $\{a_i^\mu,b_i^\mu\}$, for which we shall resort to the replica trick \cite{mpv}. Next, following Gardner \cite{gar}, we impose that the average distance between solutions vanishes (i.e. that the typical volume reduces to a single point). This condition intuitively marks $\rho^\star$, if an increase of $\rho$ reduces the set of flux vectors satisfying (\ref{vn}) until no more solutions are found. We will see that this Ansatz describes the system correctly only up to a critical value of $n$. Above this point, the solution is no longer unique. The breakdown of the Ansatz is directly linked to the existence of a conserved metabolic pool.

The volume of flux vectors satisfying (\ref{vn}) is given by
\begin{equation}
V(\rho) = \int_0^\infty d\boldsymbol{s} \prod_{\mu=1}^M
 \theta \left[ \frac{1}{\sqrt{N}}\sum_{i=1}^N s_i(a_i^\mu-\rho b_i^\mu) \right]
\end{equation}
where the extra $\sqrt{N}$ factor (here and in (\ref{avg})) is added for convenience. By analogy with known systems with quenched disorder, we expect the typical volume of solutions at fixed $\rho$ to be given by $W(\rho)\sim e^{N v(\rho)}$, where (with over-bar denoting the average over the quenched disorder)
\begin{equation}
v(\rho)=\lim_{N\to\infty}\frac{1}{N}\overline{\log V(g)}
\end{equation}
is a self-averaging quantity. Note that the disorder-averaging includes the constraints (\ref{con}), so
\begin{equation}\label{avg}
\overline{X}=\frac{\langle X\prod_{i=1}^N\delta[N^{-1/2}\sum_{\mu=1}^g z^\mu(a_i^\mu-b_i^\mu)]\rangle}{\langle \prod_{i=1}^N\delta[N^{-1/2}\sum_{\mu=1}^g z^\mu(a_i^\mu-b_i^\mu)]\rangle}
\end{equation}
where $\langle\cdots\rangle$ is now an average over the free, un-constrained stoichiometric coefficients $\{a_i^\mu,b_i^\mu\}$.  In turn, the quenched average can be computed via the replica trick:
\begin{equation}
\overline{\log V(g)}=\lim_{r\to 0}\frac{1}{r}\log \overline{V(g)^r}
\end{equation}

As in \cite{fulc}, it is convenient to write the stoichiometric coefficients as $a_i^\mu=\overline{a}(1+\xi_i^\mu)$ and $b_i^\mu=\overline{b}(1+\eta_i^\mu)$, with $\xi_i^\mu$ and $\eta_i^\mu$ zero-average Gaussian random variables. Inserting these in (\ref{vn}) one immediately sees that, to leading order in $N$, the optimal growth rate depends only on the average input and output coefficients: $\rho^\star=\overline{a}/\overline{b}$. In turn, the corrections to the leading order can be captured by re-writing $\rho$ as
\begin{equation}
\rho=\frac{\overline{a}}{\overline{b}}\left(1+\frac{g}{\sqrt{N}}\right)
\end{equation}
(\ref{con}), however, requires that the average input and output coefficients are the same (this can be easily seen by direct substitution). Hence we shall set $\rho\sim  e^{g/\sqrt{N}}$ and shift the focus to $g$: $g>0$ (resp. $g<0$) now signals an expanding (resp. contracting) phase. The calculation in the limit $N\to\infty$ can be carried out along the lines of \cite{fulc}, except that the new constraints (\ref{con}) and the fact that the average flux is now a free variable (in \cite{fulc} it was conveniently fixed to $1$ because of the invariance of (\ref{vn}) under re-scalings $s_i\to\lambda s_i~~\forall{i}$) lead to the introduction of extra order parameters. The key one is however still the overlap 
\begin{equation}
q_{\alpha\beta}=\frac{1}{N}\sum_i s_{i\alpha}s_{i\beta}
\end{equation}
between different solutions $\boldsymbol{s}_\alpha=\{s_{i\alpha}\}$ and $\boldsymbol{s}_\beta=\{s_{i\beta}\}$ (at fixed $g$). In the replica-symmetric approximation (which is putatively exact in the present case due to the convexity of the  space of solutions), 
\begin{equation}
q_{\alpha\beta}=q+\chi\delta_{\alpha\beta}
\end{equation}
and one is lead to consider the following saddle-point problem:
\begin{multline}\label{pro}
v\left( g \right) = \max_{m, q, \chi, u} \Big[ H_1 \left( m, q, \chi, u \right) +\\
+ \max_{\beta, \tau, \hat{m}, \hat{u}} H_2\left(m, \hat{m}, q, \chi, \beta, \tau, u, \hat{u} \right) \Big]
\end{multline}
where
\begin{equation}\label{h1}
H_1=\frac{1}{n}\left< \log \int \frac{dc} {\sqrt{2\pi k \chi}} e^{-\frac{{(c+gm+\xi\sqrt{kq}-ik \phi u)}^2}{2k\chi}} \right>_\xi 
\end{equation}
and
\begin{multline}\label{h2}
H_2= \hat{u}u + \hat{m}m + \frac{\beta(q+\chi)}{2} -\frac{\tau^2\chi}{2} +\\ 
  + \left< \log \int ds \: e^{-\frac{\beta s^2}{2}-s \left( \xi \sqrt{\frac{n\hat{u}^2}{k \phi} + \tau^2}+\hat{m} \right)} \right>_\xi
\end{multline}
In (\ref{h1}) and (\ref{h2}), $\langle\cdots\rangle_\xi$ denotes an average over the unit Gaussian random variable $\xi$, while $k=\overline{(\eta_i^\mu-\xi_i^\mu)^2}$ is a parameter related to the quenched disorder distribution.

\begin{figure}
\begin{center}
\includegraphics*[width=8.5cm]{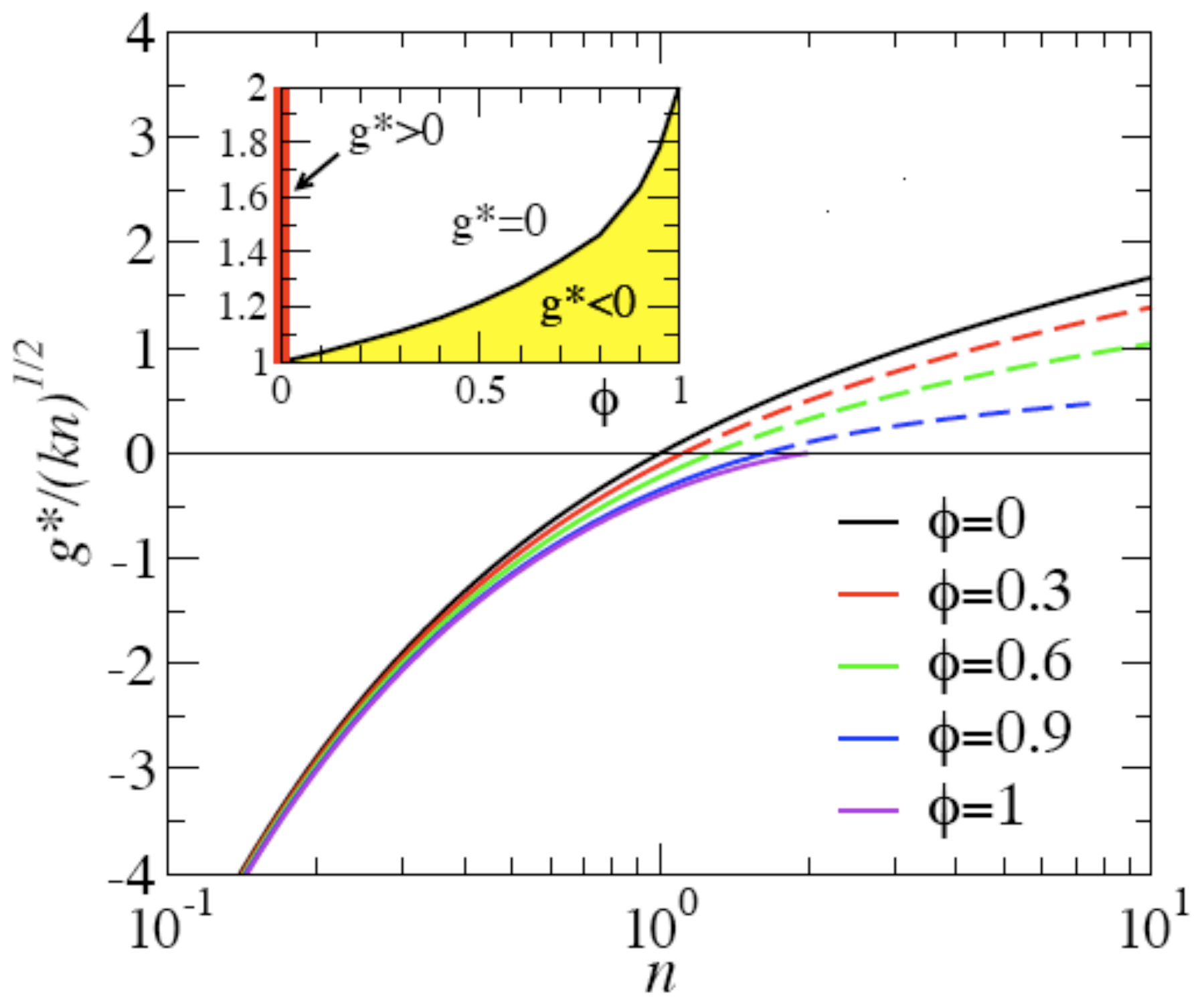}
\caption{\label{gstella}$g^\star/\sqrt{kn}$ versus $n$ for different values of $\phi$. Inset: $n_c$ versus $\phi$. For $\phi>0$, the continuous curves (corresponding to the saddle point that is valid in the region where the unique-solution anstaz holds) have been continued as dashed lines where multiple flux states satisfy (\ref{vn}) with (\ref{con}). The highlighted $y$-axis in the inset marks the region in the $(\phi,n)$ phase diagram where an expanding phase is possible with $g^\star>0$ occurs.}
\end{center}
\end{figure}

\begin{figure}
\begin{center}
\includegraphics*[width=8.5cm]{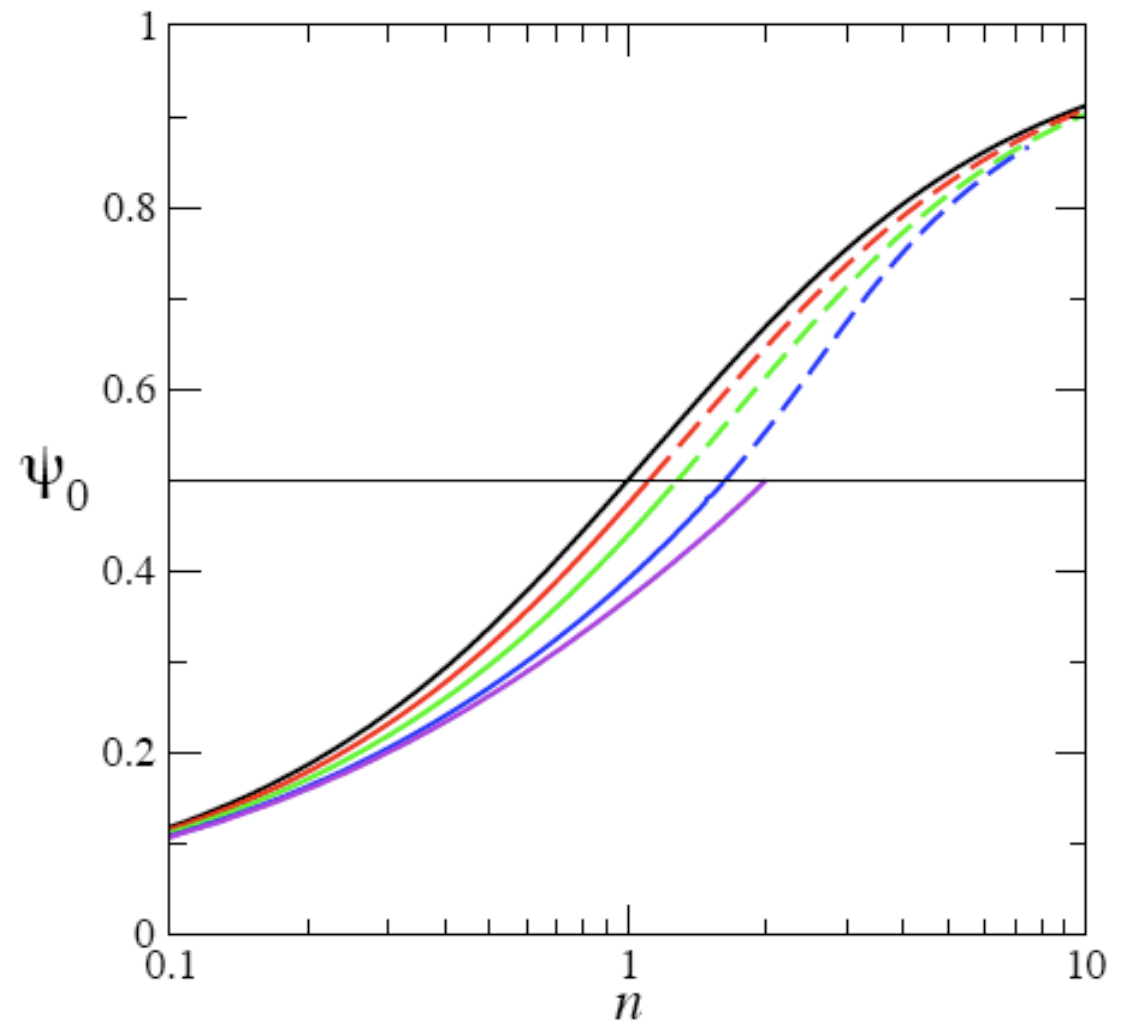}
\includegraphics*[width=8.5cm]{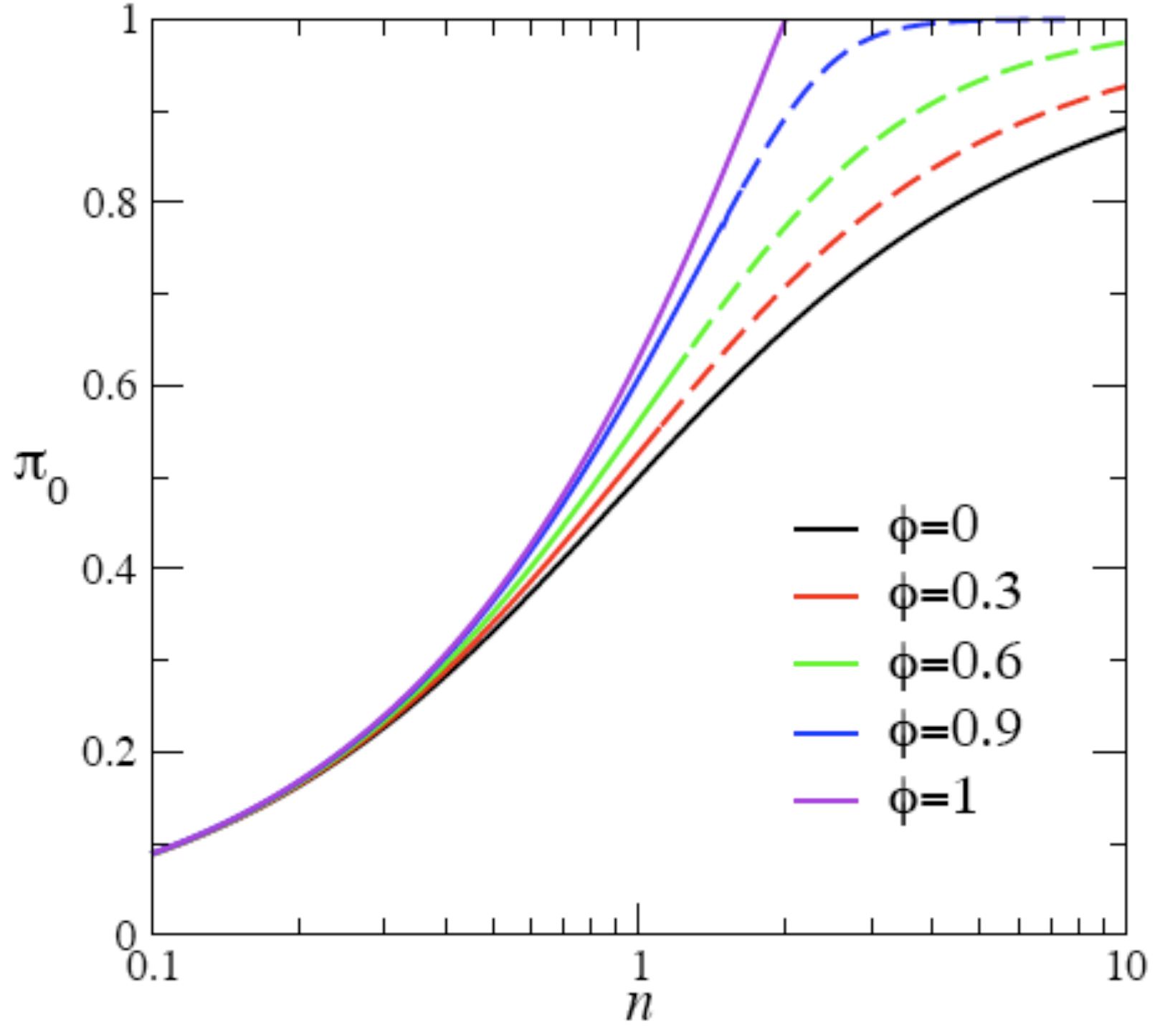}
\caption{\label{phipsi}Fraction of null reactions ($\psi_0$, top panel) and fraction of intermediate reagents ($\pi_0$, bottom panel) versus $n$ for different values of $\phi$. Continuous/dashed lines convention as in Fig.  \ref{gstella}.}
\end{center}
\end{figure}

Physically, the order parameter $\chi$ measures the distance between solutions (it is easy to see that indeed $(1/N)\sum_i(s_{i\alpha}-s_{i\beta})^2=2\chi(1-\delta_{\alpha\beta})$). It is reasonable to expect that as $g$ increases, the typical volume of solutions shrinks as it gets more and more difficult to satisfy (\ref{vn}). Assuming that a single flux state satisfies all constraints when $g=g^\star$ is then equivalent to studying the solution of (\ref{pro}) in the limit $\chi\to 0$. This can be done by introducing a proper re-scaling of the order parameters in terms of $\chi$, similarly to \cite{fulc}, so as to allow the integrals in (\ref{h1}) and (\ref{h2}) to be evaluated by steepest descent. In the present case, it is convenient to set
\begin{gather}
b = \chi \beta, ~~~~ \hat{v} = i \chi \frac{\hat{u}}{k \phi}, ~~~~ \bar{m} = \chi \hat{m}, \\ 
\sigma = \chi \sqrt{\tau^2 + \frac{n\hat{u}^2}{k\phi}}, ~~~~ v = -i k \phi u, ~~~~t = - \frac{\bar{m}}{\sigma}
\end{gather}
With these definitions, one obtains the following set of saddle point conditions:
\begin{gather}
b = - \left< \xi \left(t - \xi \right) \theta \left( t - \xi \right) \right>_\xi \nonumber \\ 
\sigma = - m \frac{ \left< \xi \left( t - \xi \right) \theta \left( t - \xi \right) \right>_\xi}{\left< \left( t - \xi \right) \theta \left( t - \xi \right) \right>_\xi} \nonumber\\
q = m^2 \frac {\left< \left( t - \xi \right)^2 \theta \left( t - \xi \right) \right>_\xi} {\left< \left( t - \xi \right) \theta \left( t - \xi \right) \right>_\xi^2} \nonumber\\
v = n \phi \hat{v} k \label{sp}\\ 
\hat{v} = \frac{1}{n}\sqrt{\frac{q}{k}}~ f_1(g^\star,m,v,q)\nonumber\\
\sigma^2 + n v \hat{v} = \frac{q}{n} ~f_2(g^\star,m,v,q)\nonumber\\
\bar{m} = \frac{g^\star}{n} \sqrt{\frac{q}{k}}~f_1(g^\star,m,v,q)\nonumber\\
\pi_0 = n \left( 1 - \psi_0 \right)\nonumber
\end{gather}
where
\begin{gather}
f_1=\left<\left( \frac{ g^{\star} m+ v}{ \sqrt{k q}} + \xi \right) \theta \left( \frac{ g^{\star} m + v}{ \sqrt{k q}} + \xi \right) \right>_\xi\\
f_2=\frac{q}{n} \left< \left( \frac{g^{\star}m + v}{ \sqrt{ k q}} + \xi \right)^2 \theta \left( \frac{g^{\star} m + v}{ \sqrt{ k q}} + \xi \right) \right>_\xi
\end{gather}
$\pi_0$ and $\psi_0$ denote, respectively, the fraction of ``intermediate'' metabolites for which the total output equals the total input and the fraction of reactions with zero flux: in particular, 
\begin{gather}\label{al}
\pi_0= \frac{1}{2}\left( 1 + \mathrm{erf} \frac{g^\star m + v}{\sqrt{2 kq}} \right) \\
\psi_0 = \frac{1}{2} \mathrm{erfc} \frac{t}{\sqrt{2}}\label{be}
\end{gather}
Equations (\ref{sp}) can be solved numerically to obtain $g^\star$ (and all order parameters) as a function of $n$. Note that the dependence of $g^\star$ on the disorder variables is embodied in the combination $g^\star/\sqrt{k}$, meaning that the optimal growth rate is larger the bigger is the spread in the difference between input and output coefficients (in other words, the system maximizes growth by taking advantage of imbalances between inputs and outputs in the stoichiometric coefficients). Results for $g^\star/\sqrt{nk}$ as a function of $n$ are shown in Fig. \ref{gstella}.
The critical point $n_c$ where $g^\star=0$ (now a function of $\phi$) becomes larger as $\phi$ increases and ultimately, when all metabolites form a conserved pool, becomes equal to two. This signals that conserved metabolic pools reduce the optimal growth capabilities, e.g. twice as many reactions are required to ensure a steady state with $g=0$ when $\phi=1$ than when $\phi=0$. Similarly, the quantities $\psi_0$ and $\pi_0$ defined in (\ref{al}) and (\ref{be}) are displayed in Fig. \ref{phipsi}.
Notice that as $\phi$ increases a larger and larger fraction of reagents is fully re-cycled into production, while the fraction of zero fluxes becomes consistently smaller as $\phi$ increases (as more active reactions are required to keep a steady growth rate). 

\begin{figure}
\begin{center}
\includegraphics*[width=8.5cm]{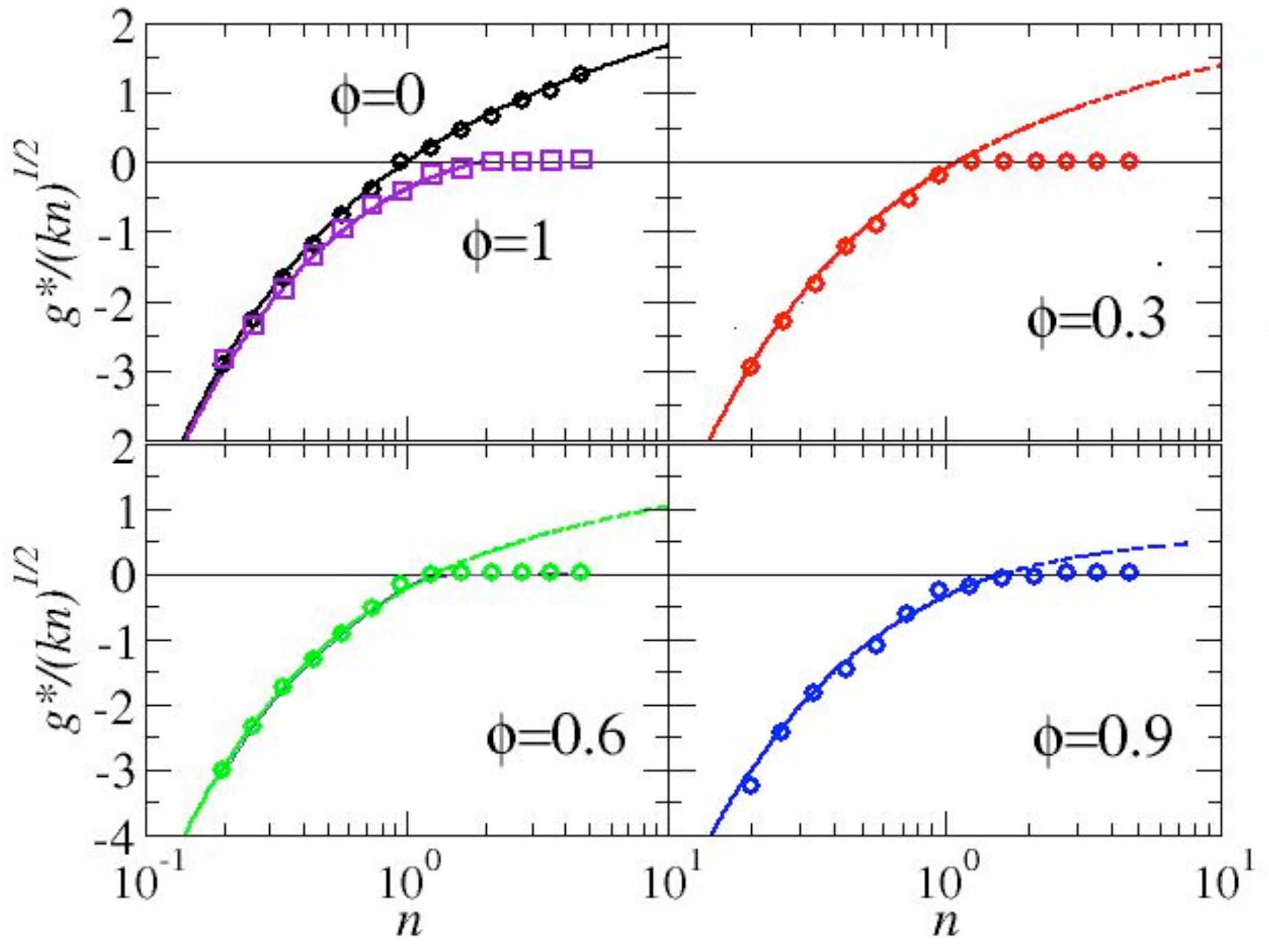}
\caption{\label{mino}Analytical (line) and numerical (markers) results for $g^\star$ versus $n$. The numerical solution has been calculated by MinOver$^+$ \cite{graph} on systems with $NM=10^4$ (variable sizes: $44\leq N\leq 216$, $46\leq M\leq 220$). Corresponding values of $\phi$ are displayed in the different panels. Averages over 50 disorder samples. The intrinsic resolution of $\rho$ used for these computer experiments is $\delta\rho=10^{-4}$.}
\end{center}
\end{figure}

\begin{figure}
\begin{center}
\includegraphics*[width=8.5cm]{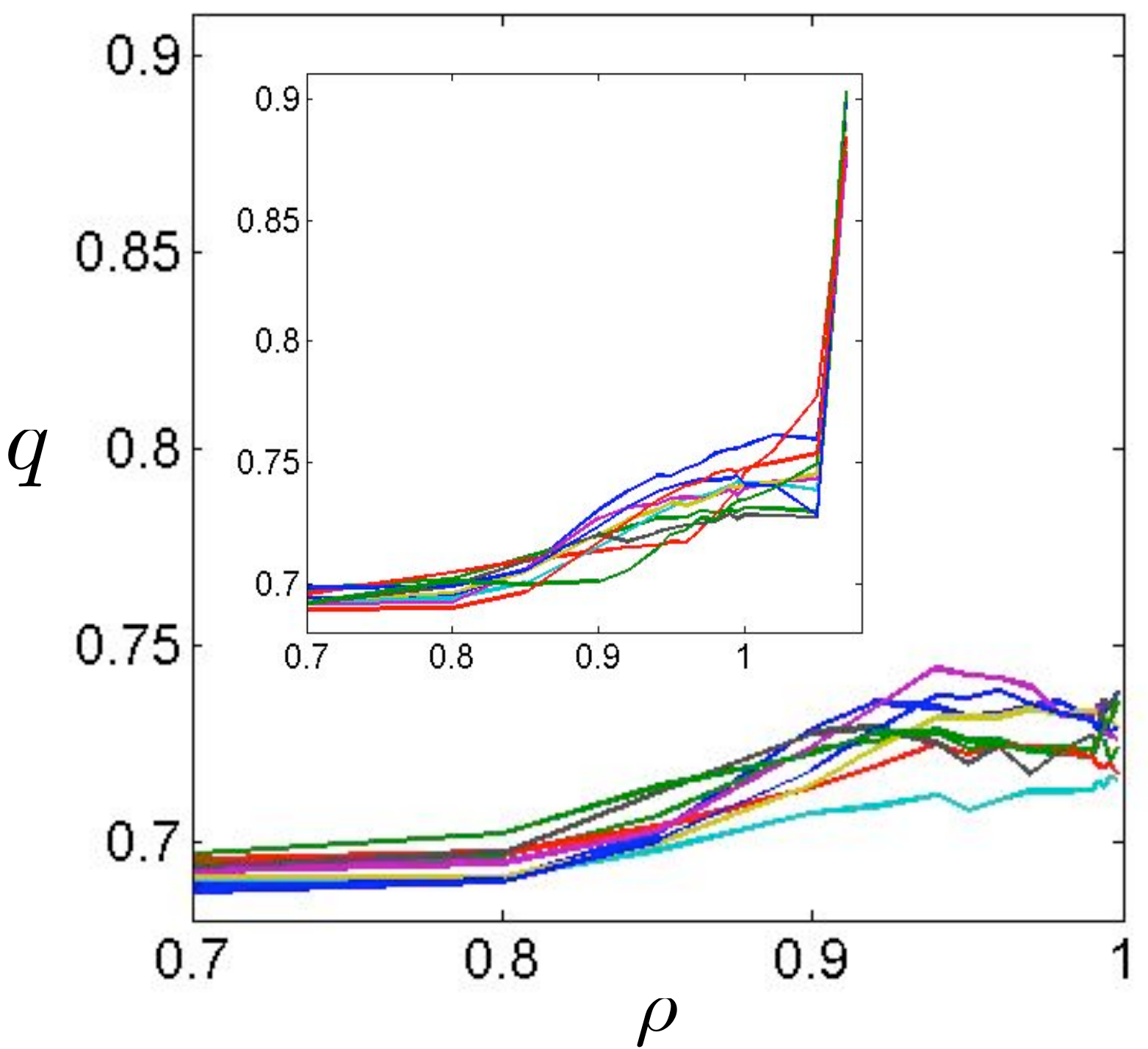}
\caption{\label{ovl}Average overlap between $100$ different solutions of (\ref{vn}) versus $\rho$ for a network with $N=100$ and $M=50$ ($n=2$) in the presence of a single conserved pool formed by 20 metabolites ($\phi=0.4$). For this system, $\rho^\star=1$. Different lines corresponding to different disorder realizations are shown, to stress sample-independence of the qualitative picture. Inset: same, without the conserved pool ($\phi=0$). For this system, $\rho^\star\simeq 1.066$.}
\end{center}
\end{figure}

These results can be confirmed numerically by calculating $g^\star$ via the MinOver$^+$ algorithm introduced in \cite{graph} based on \cite{km} (see Fig. \ref{mino}). Its main idea is to rotate the vector $\{s_i\}$ iteratively in the direction of the worst-satisfied constraint, until all constraints are satisfied at fixed $\rho$. Then one can increase $\rho$ by a quantity $\delta\rho$ and iterate the procedure, until no more solutions are found.
Simulations suggest that as $g\to g^\star$ the solution is unique for $n<n_c$ (where $g^\star<0$), while for $n>n_c$ the assumption of a unique solution that underlies the $\chi\to 0$ limit ceases to be correct and multiple solutions must exist. This can be seen directly in numerical experiments by measuring the overlap
\begin{equation}\label{over}
q_{\alpha\beta}=\frac{2}{N}\sum_{i=1}^N\frac{s_{i\alpha}s_{i\beta}}{s_{i\alpha}^2+s_{i\beta}^2}
\end{equation}
between different solutions $\boldsymbol{s}_\alpha$ and $\boldsymbol{s}_\beta$ obtained by MinOver$^+$. If $\boldsymbol{s}_\alpha=\boldsymbol{s}_\beta$, then $q_{\alpha\beta}=1$. In general, $0\leq q_{\alpha\beta}\leq 1$. (Note that the definition (\ref{over}) must be corrected to deal with null fluxes, since the overlap of the null solution with itself must obviously equal one.) In particular, we are interested in the average of $q_{\alpha\beta}$ (averaged over many solutions), denoted by $q$. This is reported in Fig. \ref{ovl}.
It is natural to expect that as $\rho$ increases, the volume of solutions decreases and correspondingly $q$ increases. We see that indeed the volume initially shrinks but it stops contracting before the optimal growth rate is reached. Ultimately, $q<1$ at $\rho^\star$, confirming that many flux states satisfy (\ref{vn}). By contrast, when no conserved pools are present $q$ keeps growing and tends to $1$ as $\rho\to\rho^\star$, as a unique solution exists.

From a biological viewpoint, one sees that stoichiometric constraints on one hand reduce the production capabilities limiting the system to a stationary state. On the other hand, they increase robustness, since many (microscopic) flux states are compatible with the optimal (macroscopic) growth performance. Clearly, increased flexibility is crucial for biological stability, since environmental changes or localized flux variations are more likely to be sustained by the system by re-arranging fluxes while keeping maximal production rates. This trade-off between stability and optimal growth has been recently observed in the metabolic network of the bacterium {\it E. coli} \cite{forth}: a finite volume of flux vectors is indeed compatible with a maximal growth assumption for {\it E. coli} and the predicted flux range is in good agreement with the measured experimental fluxes in different environments. 

\begin{figure}
\begin{center}
\includegraphics*[width=8.5cm]{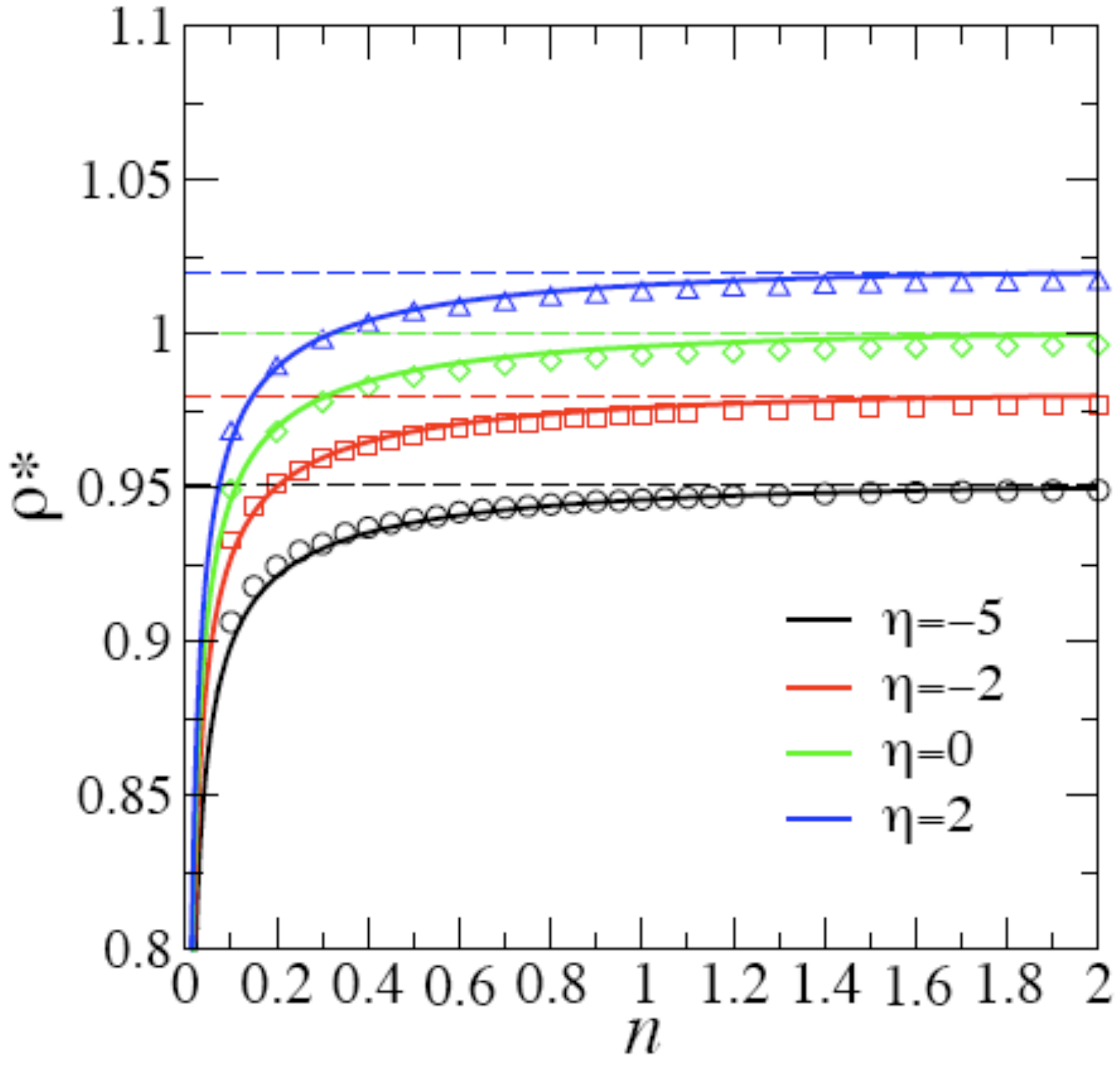}
\caption{\label{bla}Analytical (lines) and numerical (markers) values of $\rho^\star$ versus $n$ for different $\eta$. The numerical solution is for a system of $M=100$ metabolites. Analytical lines are given by $\rho(n)=\overline{\rho}_\eta(1+g^\star/\sqrt{nM})$, with $g^\star$ given by the solution of (\ref{sp}). The horizontal dashed lines are placed at $\overline{\rho}_\eta$.}
\end{center}
\end{figure}

It is easy to extend these results to the case in which the reaction network presents dispersion at reaction nodes. The simplest possibility is to consider, instead of (\ref{con}), the condition
\begin{equation}\label{coneta}
\sum_{\mu=1}^Mz^\mu(a_i^\mu-b_i^\mu)=\eta_i
\end{equation}
with $\eta_i$'s real constants. A positive (respectively, negative) $\eta_i$ says that for reaction $i$ more (respectively, less) reagents are produced than they are consumed. While the typical properties of large, random instances with random $\eta_i$'s can be characterized analytically along the lines described above, it is more useful to concentrate on the case where $\eta_i=\eta~~\forall i$, which provides for an immediate interpretation of the limiting cases $\eta>0$ and $\eta<0$. It is easy to understand that (\ref{coneta}) implies that the average input and output stoichiometric coefficients must be linked by
\begin{equation}
\overline{a}-\overline{b}=\frac{\eta}{\phi M}
\end{equation}
This in turn yields 
\begin{equation}
\rho^\star\leq \overline{\rho}_\eta\equiv\left(1-\frac{\eta}{\phi\overline{a}M}\right)^{-1}
\end{equation}
Hence, the influence of $\eta$ is essentially to allow, in a finite system, for an expanding phase if $\eta>0$. If $\eta<0$, instead, the system is necessarily confined to a  contracting regime. In particular, the introduction of $\eta$ does not affect $g^\star$. Analytical and numerical studies fully confirm this prediction (see Fig. \ref{bla}).

To summarize, the existence of conserved metabolic pools (groups of reagents whose aggregate concentration is invariant in time) stems directly from stoichiometric balance and characterizes all biochemical reaction networks. The model we addressed aims at understanding how these invariant structures affect the global growth properties within a simple exactly solvable setting. We have seen that stoichiometry frustrates the system by reducing its optimal production capability. At the same time, a finite volume of flux configurations is compatible with the optimal conditions, implying an increase of stability and flexibility. Both ingredients are crucial at the biological level since, reasonably, metabolic networks should be robust to local flux variations. The basic mechanism by which local conservation laws build up to regulate the trade-off between growth and stability in the present model is likely to lie at the core of recently observed properties of real metabolic networks. An important aspect that cannot be analyzed theoretically at the level of a fully-connected model is how dynamical invariants affect the way in which reaction networks re-arrange fluxes in response to knock-outs or environmental changes. It is reasonable to expect that the graphical version of the problem, introduced in \cite{graph}, will be useful to shed some light on this issue.

\acknowledgments

We are indebted with S. Franz and T. J\"org for useful comments and questions.

\end{document}